\RequirePackage{fix-cm}
\documentclass[smallextended]{svjour3}       
\smartqed  
\usepackage{graphicx}
 \journalname{myjournal}
\usepackage{dcolumn}
\usepackage{bm}
\usepackage{braket}
\usepackage{amsmath}
\usepackage{amsfonts}
\usepackage{amssymb}
\usepackage{xcolor}
\usepackage{hyperref}
\newcommand{\rmd}{\mathrm{d}}

\newcommand{\vn}{\mathbf{n}}
\graphicspath{{./figures/}}
\begin{document}

\title{Tomographic entanglement indicators in multipartite systems}

\author{B.~Sharmila$^{1}$\and
        S.~Lakshmibala$^{1}$\and
        V.~Balakrishnan$^{1}$
}

\institute{B. Sharmila \at
              \email{sharmilab@physics.iitm.ac.in}           
           \and
           $\phantom{x}^{1}$ \hspace*{1em} Department of Physics, Indian Institute of Technology Madras, Chennai 600036, India
}

\date{Received: date / Accepted: date}

\maketitle

\begin{abstract}
We assess the performance of an entanglement indicator which can be obtained directly from tomograms,  avoiding state reconstruction procedures. In earlier work, 
we have examined this tomographic entanglement indicator, and a variant obtained from it, in the context of continuous variable systems. It has been shown that, in multipartite systems of radiation fields, these indicators fare as well as standard measures of entanglement. In this paper we assess these indicators in the case of two generic {\it hybrid} quantum systems,  the double Jaynes-Cummings model and the double Tavis-Cummings model using, for purposes of comparison, the quantum mutual information as a standard reference for both quantum correlations and entanglement. The dynamics of entanglement is investigated in both models over a sufficiently long time interval. We establish that the tomographic indicator provides a good estimate of the extent of entanglement both in the atomic subsystems and in the field subsystems. 
An indicator obtained from the tomographic 
indicator  as an approximation, however,  does not capture the entanglement properties of  atomic subsystems, although it is useful for field subsystems. Our results are inferred from numerical calculations based on the two models, simulations of relevant equivalent circuits in both cases, and experiments performed on the IBM computing platform.

\end{abstract}


\section{Introduction}
\label{intro}
Several interesting effects are observed through entanglement dynamics in models of hybrid quantum systems where spins  
are coupled to  continuous dynamical variables. 
Among other possibilities, these  models  also describe  atoms interacting with radiation fields. Interesting phenomena such as sudden death and birth of entanglement are seen in the double Jaynes-Cummings (JC) model~\cite{eberly}  and the double Tavis-Cummings (TC) model~\cite{dtcm}, both 
of which  have been examined extensively in 
theory and experiment~\cite{jcm1,jcm2,tcm1}. 
Furthermore, a collapse 
of the entanglement to a constant non-zero value 
over a significant interval of time occurs in  
tripartite models  of a ${\rm \Lambda}$-atom interacting with two radiation fields~\cite{pradip1}, or in an optomechanical set-up where the radiation field interacts with an atom and a mechanical oscillator~\cite{pradip3}.
 
It is evident that, in these investigations of 
entanglement dynamics, it is necessary to 
identify  appropriate quantifiers of entanglement 
at every instant of time. Quantifiers used extensively, 
such as the subsystem von Neumann entropy  
$\xi_{\textsc{svne}}$ and the subsystem linear entropy 
$\xi_{\textsc{sle}}$, are obtained  from the 
reduced density matrix $\rho$ corresponding to the 
subsystem of interest according to  
$\xi_{\textsc{svne}}=-\mathrm{Tr}\,(\rho \,\log \,\rho)$ 
and $\xi_{\textsc{sle}}=1-\mathrm{Tr}\,(\rho^{2})$.  
Reconstructing the density matrix  
 from experimental data which are typically in the form of tomograms (or, equivalently,  quadrature histograms), 
 however,  is an elaborate and tedious statistical procedure that  
 is inherently error-prone.
It is therefore desirable to extract information 
about the state {\em  directly} from the tomograms, avoiding explicit state reconstruction. In bipartite qubit systems, the efficacy of such a program has been assessed by  estimating relevant nonlinear functions of the density matrix directly from the tomogram (see, for instance,~\cite{HuiKhoon}). In the context of continuous variable systems, a {\em qualitative} indicator 
 of entanglement using tomograms has been proposed in Ref.~\cite{sudhrohithbs}. 

In earlier work~\cite{sharmila}, we   
identified  a tomographic entanglement indicator 
$\xi_{\textsc{tei}}$ that quantifies the extent of entanglement directly from the relevant tomograms,  
and assessed its performance vis-\`a-vis 
$\xi_{\textsc{svne}}$ and 
$\xi_{\textsc{sle}}$ in a double-well BEC system with inherent nonlinearities.  We also carried out  a comparative study between $\xi_{\textsc{tei}}$ and an entanglement 
indicator $\xi_{\textsc{ipr}}$ obtained from the inverse participation ratio both in the BEC system  and in a nonlinear model of a multi-level atom interacting with a radiation field. This investigation brings into focus the role of the initial state considered and the nature of the nonlinearity in the model system, in determining the performance of $\xi_{\textsc{tei}}$ and $\xi_{\textsc{ipr}}$ \cite{sharmila2} as entanglement 
indicators. 
We note that both the systems considered for 
our purposes are bipartite in nature, with the subsystems modelled as oscillators. In this paper,  we extend our investigations on the tomographic entanglement indicator to multipartite {\em hybrid} quantum systems. 

 A good measure of the entanglement between any two subsystems $A$ and $B$ of a multipartite system is the quantum mutual information,  defined as 
 \begin{equation}
 \label{eqn:qmi}
 \xi_{\textsc{qmi}}^{(AB)}=\xi_{\textsc{svne}}^{(A)}+\xi_{\textsc{svne}}^{(B)}-\xi_{\textsc{svne}}^{(AB)}.
 \end{equation}
 The terms on the right-hand side are, respectively, 
  the subsystem von Neumann entropies of $A$, $B$ and the bipartite subsystem $AB$. 
In this paper, we compare the performance of 
$\xi_{\textsc{tei}}$ as a measure of entanglement 
with that of  $\xi_{\textsc{qmi}}$ during dynamical 
evolution  in the double JC and the double TC models. In view of the extensive work being carried out currently in constructing quantum circuits for various models of quantum 
optics~\cite{circuits}, we have  also constructed a quantum circuit to mimic the dynamics  of the double JC model using the IBM computing platform,  and obtained the tomogram at a specific instant of time. From this we have computed 
$\xi_{\textsc{tei}}$ at that instant. We have also substantiated our results by numerically simulating the dynamics of both the model and the equivalent circuit. For the latter, we have used the IBM Open quantum assembly language (QASM) simulator~\cite{ibm_main,qiskit}.

The plan of the rest of this paper is as follows.  In Sec. 
\ref{sec:2}, we outline the procedures used to obtain the relevant entanglement measures. In Sec. \ref{sec:models}, we describe the two hybrid models mentioned above, and compare the performance of the various measures during time evolution. 
We further construct and examine the equivalent circuit for the double JC model, extract the indicators, and compare them with those from numerical simulation.  Similar procedures have been carried out for the double TC model, and conclusions have been drawn based on the experiment, simulation and numerical analysis.

\section{Entanglement indicators from tomograms}
\label{sec:2}
We start with a brief  
review of  the procedure for obtaining 
$\xi_{\textsc{tei}}$.  Of immediate relevance to us  are the optical tomogram corresponding to the radiation field and the spin tomogram of an  atom, at any instant of time. A 
tomogram is a histogram of experimental outcomes of 
the measurement of an appropriate set of observables; the latter are judiciously selected to yield maximal information about the quantum state.
In the case of  a single-mode radiation field, this is  the set of rotated quadrature operators  \cite{VogelRisken,ibort} 
 \begin{equation}
\label{eqn:quadop}
\mathbb{X}_{\theta} 
= (a^\dagger \,e^{i \theta} + a \,e^{-i \theta})/\sqrt{2}
\end{equation}
where $0 \leq \theta <  \pi$, and $a$ and $a^{\dagger}$ are photon annihilation and creation operators satisfying $[a,a^{\dagger}]=1$. The eigenvalue equation 
for the operator $\mathbb{X}_{\theta}$ is, in an obvious 
notation, $\mathbb{X}_{\theta}\ket{X_{\theta},\theta}=X_{\theta}\ket{X_{\theta},\theta}$. The expectation value of the field density matrix $\rho_{\textsc{f}}$ can be computed in each complete basis set $\lbrace\ket{X_{\theta},\theta}\rbrace$ for a given value of $\theta$. The optical tomogram \cite{VogelRisken,LvovskyRaymer} is then 
given by 
\begin{equation}
\label{optholodefn}
w(X_\theta, \theta) = \bra{X_\theta, \theta} \rho_{\textsc{f}} \ket{X_\theta, \theta}.
\end{equation}

For an atomic qubit with ground state  $\ket{g}$ and 
excited state $\ket{e}$, the set of observables is given by 
the operators 
\begin{equation}
\label{eqn:atomops}
\sigma_{x}=\tfrac{1}{2} (\ket{e}\bra{g}+\ket{g}\bra{e}), 
\;\sigma_{y}=\tfrac{1}{2} i(\ket{g}\bra{e}-\ket{e}\bra{g}), 
\;\sigma_{z}=\tfrac{1}{2} (\ket{e}\bra{e}-\ket{g}\bra{g}).
\end{equation}
 These observables yield maximal information about the atomic states~\cite{thew}. Let $\sigma_{z}\ket{m}=m\ket{m}$. Then $U(\vartheta,\varphi) \ket{m} =  \ket{\vartheta,\varphi,m}$, where 
 $U(\vartheta,\varphi) \equiv  U(\vn)$ is a general SU(2) transformation
 parametrized by the  
 polar and azimuthal angles 
 $(\vartheta, \varphi)$, or, equivalently, by the unit vector 
 $\vn$.  
 The spin tomogram is given by 
 \begin{equation}
 \label{eqn:spintomogram}
 w(\vn,m)=\bra{\vn,m}\rho_{\textsc{s}}\ket{\vn,m}
 \end{equation}
  where $\rho_{\textsc{s}}$ is the spin density matrix. 
 Different values  of $\vartheta$ and  $\varphi$ 
 give different complete  basis sets.
 
An extension of the foregoing 
to multipartite tomograms is straightforward~\cite{ibort}. 
The tomogram corresponding to a system comprising two radiation fields $A$ and $B$ and two atoms $C$ and $D$
is the (diagonal) matrix element of
the density matrix $\rho_{\textsc{abcd}}$ 
of the full system in the state 
%\begin{align}
%\nonumber &\ket{X_{\theta_{\textsc{a}}},\theta_{\textsc{a}}; X_{\theta_{\textsc{b}}},\theta_{\textsc{b}};\vn_{\textsc{c}}, m_{\textsc{c}};\vn_{\textsc{d}}, m_{\textsc{d}}} \equiv \\
%& \qquad \ket{X_{\theta_{\textsc{a}}},\theta_{\textsc{a}}} \otimes \ket{X_{\theta_{\textsc{b}}},\theta_{\textsc{b}}} \otimes \ket{\vn_{\textsc{c}}, m_{\textsc{c}}}\otimes \ket{\vn_{\textsc{d}}, m_{\textsc{d}}}.
%\end{align}
\begin{equation}
\ket{X_{\theta_{\textsc{a}}},\theta_{\textsc{a}}; X_{\theta_{\textsc{b}}},\theta_{\textsc{b}};\vn_{\textsc{c}}, m_{\textsc{c}};\vn_{\textsc{d}}, m_{\textsc{d}}} \equiv 
 \ket{X_{\theta_{\textsc{a}}},\theta_{\textsc{a}}} \otimes \ket{X_{\theta_{\textsc{b}}},\theta_{\textsc{b}}} \otimes \ket{\vn_{\textsc{c}}, m_{\textsc{c}}}\otimes \ket{\vn_{\textsc{d}}, m_{\textsc{d}}}
\end{equation}
in an obvious notation. 
%\begin{align}
%\nonumber w( & X_{\theta_{\textsc{a}}},\theta_{\textsc{a}}; X_{\theta_{\textsc{b}}},\theta_{\textsc{b}}; \vn_{\textsc{c}}, m_{\textsc{c}}; \vn_{\textsc{d}}, m_{\textsc{d}}) =\\
%& \bra{X_{\theta_{\textsc{a}}},\theta_{\textsc{a}};X_{\theta_{\textsc{b}}},\theta_{\textsc{b}}; \vn_{\textsc{c}}, m_{\textsc{c}};\vn_{\textsc{d}}, m_{\textsc{d}}} \rho_{\textsc{abcd}} \ket{X_{\theta_{\textsc{a}}},\theta_{\textsc{a}};X_{\theta_{\textsc{b}}},\theta_{\textsc{b}};\vn_{\textsc{c}}, m_{\textsc{c}};\vn_{\textsc{d}}, m_{\textsc{d}}}, 
%\label{eqn:tomo_defn}
%\end{align}
%The basis states of the radiation fields $A$ and $B$ are denoted by $\lbrace\ket{X_{\theta_{i}},\theta_{i}}\rbrace$ ($i=A,B$ respectively) and $\lbrace\ket{\vn_{i}, m_{i}}\rbrace$ ($i=C,D$) correspond to the atoms $C$ and $D$ respectively.
The reduced tomogram for a specific subsystem is obtained by tracing over the basis states of the other subsystems. For instance, the reduced tomogram corresponding to $A$ is
\begin{align}
\nonumber w_{A}(X_{\theta_{\textsc{a}}},\theta_{\textsc{a}}) & = \sum_{m_{\textsc{d}}}  \sum_{m_{\textsc{c}}} \int_{-\infty}^{\infty} \rmd X_{\theta_{\textsc{b}}} w(X_{\theta_{\textsc{a}}},\theta_{\textsc{a}};X_{\theta_{\textsc{b}}},\theta_{\textsc{b}}; \vn_{\textsc{c}}, m_{\textsc{c}}; \vn_{\textsc{d}}, m_{\textsc{d}})\\[4pt]
& = \bra{X_{\theta_{\textsc{a}}},\theta_{\textsc{a}}} \rho_{A} \ket{X_{\theta_{\textsc{a}}},\theta_{\textsc{a}}},
\label{eqn:red_tomo_1}
\end{align}
where $\rho_{A}$ is the reduced density matrix of the subsystem $A$. 

The extent of entanglement between any two subsystems, say $C$ and $D$ in the example above, can be estimated from the tomogram by computing the tomographic entanglement indicator $\xi_{\textsc{tei}}^{(CD)}$, as follows. The 
 two-mode tomographic entropy is given by 
%\begin{align}
%S(\vn_{\textsc{c}},\vn_{\textsc{d}}) = 
%- &\sum_{m_{\textsc{d}}}  \sum_{m_{\textsc{c}}} w_{CD}(\vn_{\textsc{c}}, m_{\textsc{c}}; \vn_{\textsc{d}}, m_{\textsc{d}})\times \nonumber\\[4pt] 
%& \log \,[w(\vn_{\textsc{c}}, m_{\textsc{c}}; \vn_{\textsc{d}}, m_{\textsc{d}})].
%\label{eqn:2_mode_entropy}
%\end{align}
\begin{equation}
S(\vn_{\textsc{c}},\vn_{\textsc{d}}) = 
- \sum_{m_{\textsc{d}}}  \sum_{m_{\textsc{c}}} w_{CD}(\vn_{\textsc{c}}, m_{\textsc{c}}; \vn_{\textsc{d}}, m_{\textsc{d}})  \log \,w_{CD}(\vn_{\textsc{c}}, m_{\textsc{c}}; \vn_{\textsc{d}}, m_{\textsc{d}}),
\label{eqn:2_mode_entropy}
\end{equation}
while the single-mode subsystem 
tomographic entropy is 
%\begin{align}
%S(\vn_{j}) = - & \sum_{m_{j}} w_{j}(\vn_{j},m_{j}) \times \nonumber \\[4pt]
%&\log \,[w_{j}(\vn_{j},m_{j})]  \; (j = C, D).
%\label{eqn:1_mode_entropy}
%\end{align}
\begin{equation}
S(\vn_{j}) = -  \sum_{m_{j}} w_{j}(\vn_{j},m_{j})\,
\log \,w_{j}(\vn_{j},m_{j})  \;\;(j = C, D).
\label{eqn:1_mode_entropy}
\end{equation}
The tomograms $w_{CD}$ and  $ w_{j}$ in 
Eqs. (\ref{eqn:2_mode_entropy}) and (\ref{eqn:1_mode_entropy}) are defined in a manner analogous to Eq. (\ref{eqn:red_tomo_1}). 
The mutual information corresponding to the 
two subsystems $C$ and $D$ is  given by  
\begin{equation}
S(\vn_{\textsc{c}};\vn_{\textsc{d}}) 
\equiv S(\vn_{\textsc{c}},\vn_{\textsc{d}}) - S(\vn_{\textsc{c}}) - S(\vn_{\textsc{d}}).
\label{mutinfo}
\end{equation}  
The tomographic entanglement indicator is then given by 
\begin{equation}
\xi_{\textsc{tei}}^{(CD)} = \langle S(\vn_{\textsc{c}};\vn_{\textsc{d}}) \rangle, 
\label{xiteidefn}
\end{equation} 
where the average is taken over the range of values of 
$\vn_{\textsc{c}}$ and $\vn_{\textsc{d}}$. Numerical 
evidence shows that a set of three orthogonal 
$\vn_{j}$'s (for each $j=C,D$) suffices 
 to obtain a  $\xi_{\textsc{tei}}^{(CD)}$ which agrees 
 reasonably well with  standard entanglement measures 
 (such as the SVNE). 
For instance, choosing  $\vn_{j}$ to be 
$(1,0,0), (0,1,0)$ and $(0,0,1)$ in turn would correspond, respectively, 
 to the eigenbasis of $\sigma_{jx}$, $\sigma_{jy}$ and 
 $\sigma_{jz}$ ($j=C,D$). It may be noted  that choosing 
 three  orthogonal $\vn_{j}$'s  is equivalent to choosing 
 three  mutually unbiased basis sets for the subsystem 
 concerned.   

Likewise, the measure $\xi_{\textsc{tei}}^{(AB)}$ 
of the 
entanglement between the fields $A$ and $B$ 
is obtained \cite{sharmila}  by averaging the corresponding mutual information
over a sufficient number of basis sets  
in the ranges $0\leqslant 
 \theta_{\textsc{a}}<\pi$ and 
 $0\leqslant \theta_{\textsc{b}}<\pi$.

\section{Entanglement indicators in hybrid multipartite models}
\label{sec:models} We now examine the indicator 
$\xi_{\textsc{tei}}$ 
 in the case of two hybrid multipartite models, namely, the double JC model \cite{eberly} and the double TC model \cite{dtcm}.

\subsection{The double Jaynes-Cummings model}

The model comprises two $2$-level atoms  $C$ and $D$ which are initially in an entangled state, with each atom interacting  with  strength 
$g$ with radiation fields $A$ and $B$ respectively. The effective Hamiltonian (setting $\hbar=1$) is \cite{eberly}  
\begin{align}
\nonumber H_{\textsc{DJC}}= &\sum_{j=A,B} 
\omega a_{j}^{\dagger} a_{j} + \tfrac{1}{2}\sum_{k=C,D} \omega_{0} \sigma_{k z} + g \,(a_{\textsc{a}}^{\dagger} \sigma_{\textsc{c} -} + a_{\textsc{a}} \sigma_{\textsc{c} +})\\
&+ g \,(a_{\textsc{b}}^{\dagger} \sigma_{\textsc{d} -} + a_{\textsc{b}} \sigma_{\textsc{d} +}).
\label{eqn:HDJC}
\end{align}
$a_{j}, a_{j}^{\dagger}$ ($j=A,B$) are photon  annihilation and creation operators, $\omega$ is the frequency of the fields, 
and $\omega_{0}$ is the  energy  
difference between the two atomic levels.
In terms of the Pauli matrices, the atomic ladder operators  
are $\sigma_{k \pm}=(\sigma_{k x} \pm i \sigma_{k y})$ 
($k = C, D$). The initial atomic states considered both in the double JC model and the double TC model are  of the form
\begin{equation}
 \ket{\psi_{0}}=\big(\ket{g}_{1}\otimes\ket{g}_{2} + \ket{e}_{1}\otimes\ket{e}_{2}\big)/\sqrt{2}
 \label{eqn:psi0_defn}
 \end{equation} 
and 
\begin{equation}
\ket{\phi_{0}}=
\big(\ket{g}_{1}\otimes\ket{e}_{2} + \ket{e}_{1}\otimes\ket{g}_{2}\big)/\sqrt{2}.
 \label{eqn:phi0_defn}
\end{equation}
Here $\ket{g}_{k}$ and  $\ket{e}_{k}$ ($k=1,2$) 
denote the respective ground and excited states 
of atom $k$.  In  the double JC model,  $1$ and $2$ are to be replaced by $C$ and  $D$ respectively.  $A$ and $B$  are initially in the zero-photon states $\ket{0}_{\textsc{a}}$ and $\ket{0}_{\textsc{b}}$. The two initial states of the full system that we consider are $\ket{0}_{\textsc{a}} \otimes \ket{0}_{\textsc{b}} \otimes \ket{\psi_{0}}_{\textsc{cd}} \equiv \ket{0;0;\psi_{0}}$ and 
$\ket{0}_{\textsc{a}} \otimes \ket{0}_{\textsc{b}} \otimes \ket{\phi_{0}}_{\textsc{cd}} \equiv \ket{0;0;\phi_{0}}$.

For  these initial states, 
we have numerically generated tomograms 
at approximately 300 instants of time, 
separated by a time step equal to $0.02$ (in units of 
$\pi/g$).  From these, we have obtained $\xi_{\textsc{tei}}$ at different instants as the system evolves.  For radiation fields,  
fairly good agreement had been 
demonstrated ~\cite{sharmila2} between  
$\xi_{\textsc{tei}}$  calculated using the procedure 
outlined earlier,  and an approximate entanglement 
indicator $\xi'_{\textsc{tei}}$ obtained by averaging 
{\em only} over the dominant values of 
$S(\theta_{\textsc{a}}:\theta_{\textsc{b}})$ 
(i.e., over a subset of values that exceed the 
mean value by more than one standard deviation).  
We now proceed to investigate if the latter 
approximation suffices even in the case of 
hybrid quantum  systems.

\begin{figure}[h]
\centering
\includegraphics[width=0.32\textwidth]{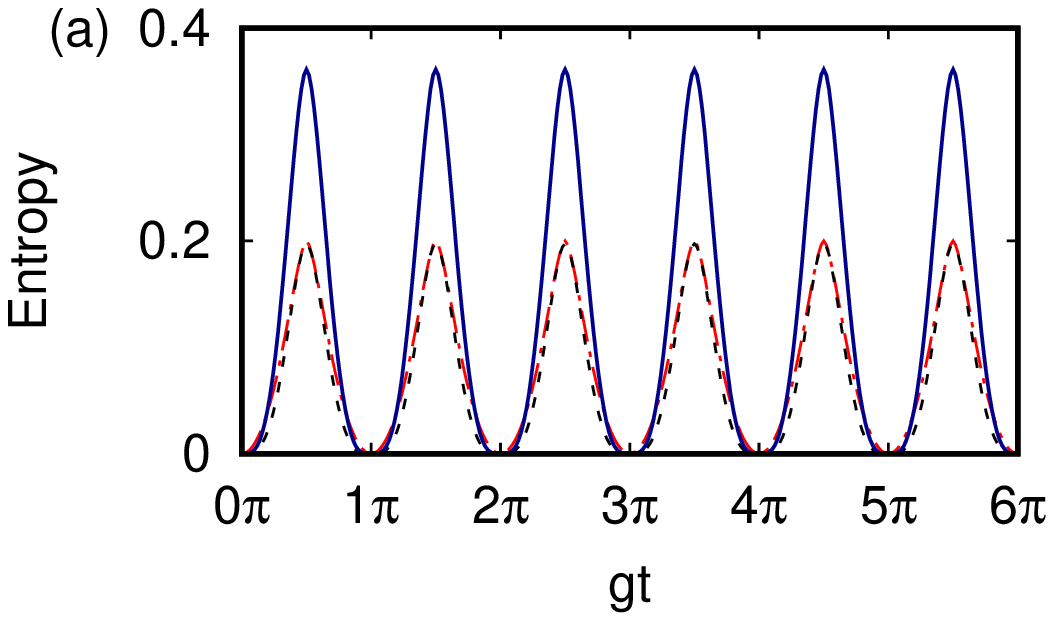}
\includegraphics[width=0.32\textwidth]{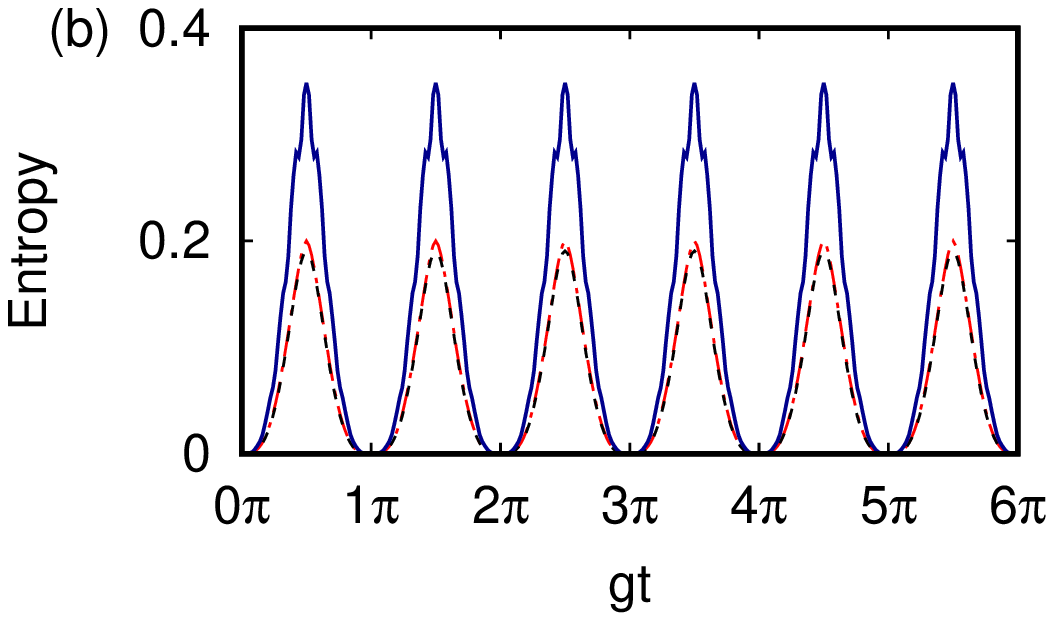}
\includegraphics[width=0.32\textwidth]{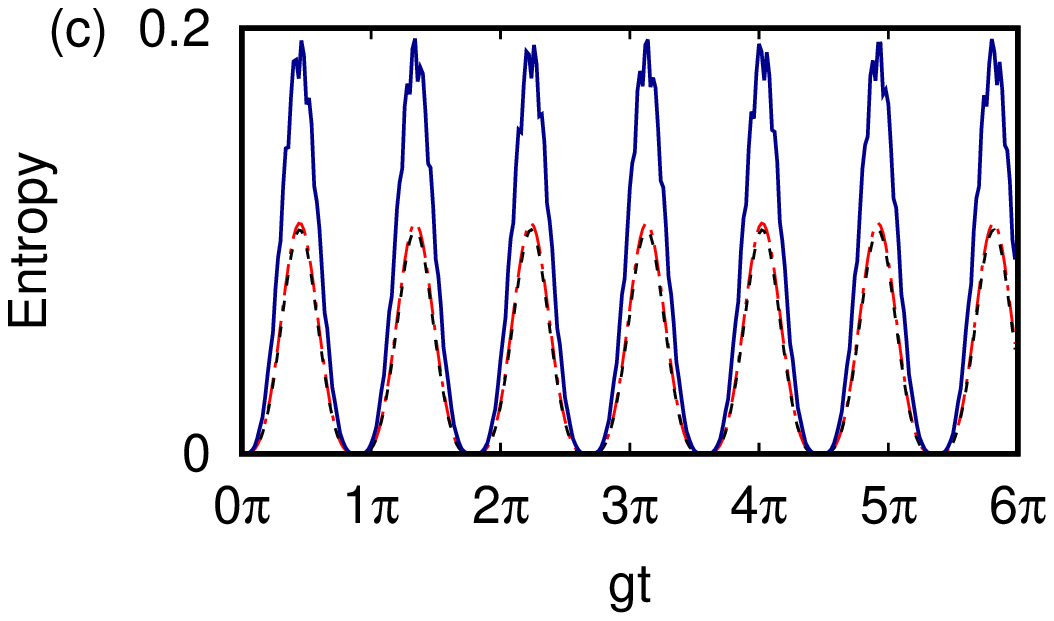}
\caption{$\xi_{\textsc{tei}}$ (black dashed curve), $\xi'_{\textsc{tei}}$ (blue solid curve) and $0.1\,\xi_{\textsc{qmi}}$ (red dot-dashed curve) 
versus scaled time $gt$  for the field subsystem in the double JC model.  (a) $\Delta=0$, initial state   $\ket{0;0;\phi_{0}}$\,
 (b) $\Delta = 0$, initial state $\ket{0;0;\psi_{0}}$\, 
 (c) $\Delta=1$, initial state $\ket{0;0;\psi_{0}}$.}
\label{fig:comp_Neq1_field}
\end{figure}

 These two entanglement indicators and 
the standard indicator  $\xi_{\textsc{qmi}}$  are 
plotted against  the scaled time $g t$ in Figs. 
\ref{fig:comp_Neq1_field} (a)-(c) 
in the case of  the field subsystems in the double 
JC model. The detuning 
parameter $ \Delta = (\omega - \omega_{0})$ has been  
set equal to zero in Figs.  
\ref{fig:comp_Neq1_field} (a) and (b),  
and to unity in Fig. \ref{fig:comp_Neq1_field} (c). 
The initial states considered are  $\ket{0;0; \phi_{0}}$ 
in Fig. \ref{fig:comp_Neq1_field} (a)
and $\ket{0;0;\psi_{0}}$ in Figs. 
\ref{fig:comp_Neq1_field} (b),(c). 
For ease of comparison, $\xi_{\textsc{qmi}}$ has been  
scaled down by a factor of $10$.  
It is evident from the figures that in this case, too, 
$\xi'_{\textsc{tei}}$ is a good approximation 
to $\xi_{\textsc{tei}}$. Both the indicators mimic 
$\xi_{\textsc{qmi}}$ closely in all the three cases 
considered. Sensitivity to the precise initial atomic state 
considered  and to the extent of detuning is revealed 
by examining the qualitative features of the indicators in the neighbourhood of their maximum values.

\begin{figure}[h]
\centering
\includegraphics[width=0.32\textwidth]{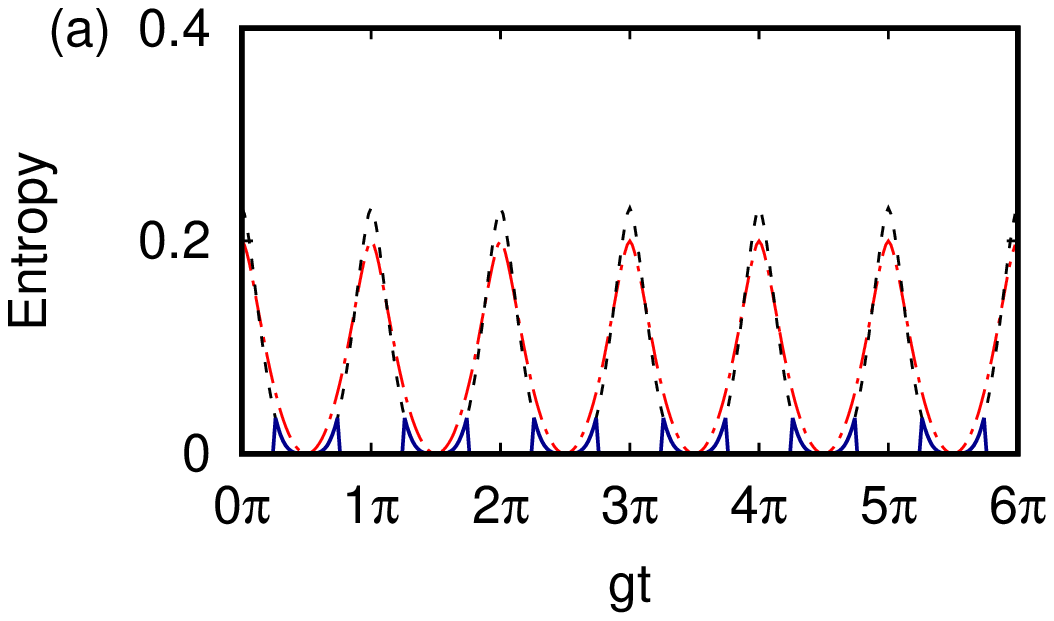}
\includegraphics[width=0.32\textwidth]{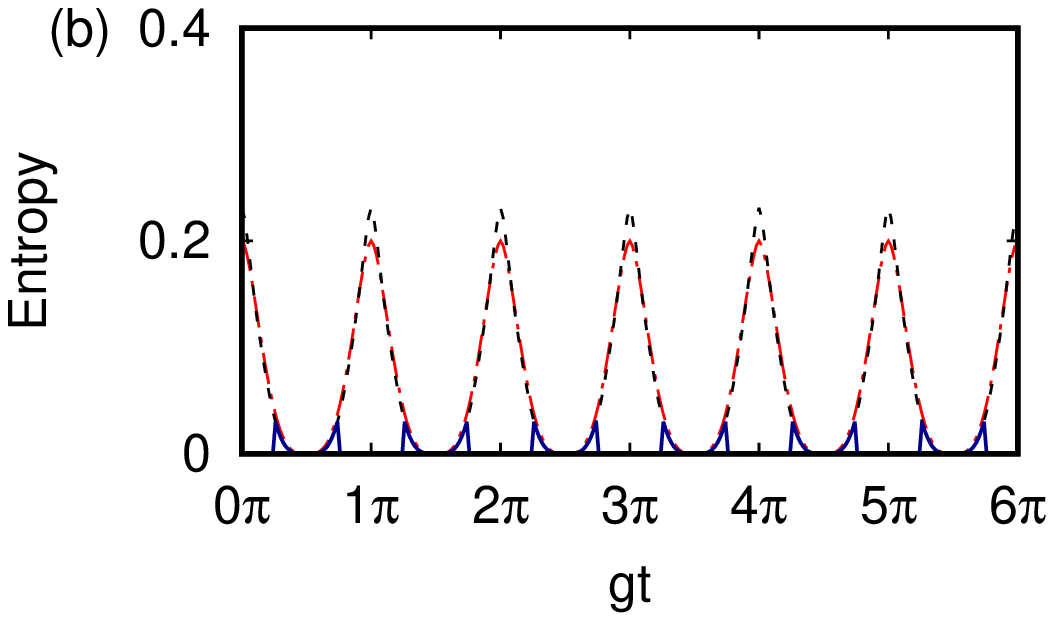}
\includegraphics[width=0.32\textwidth]{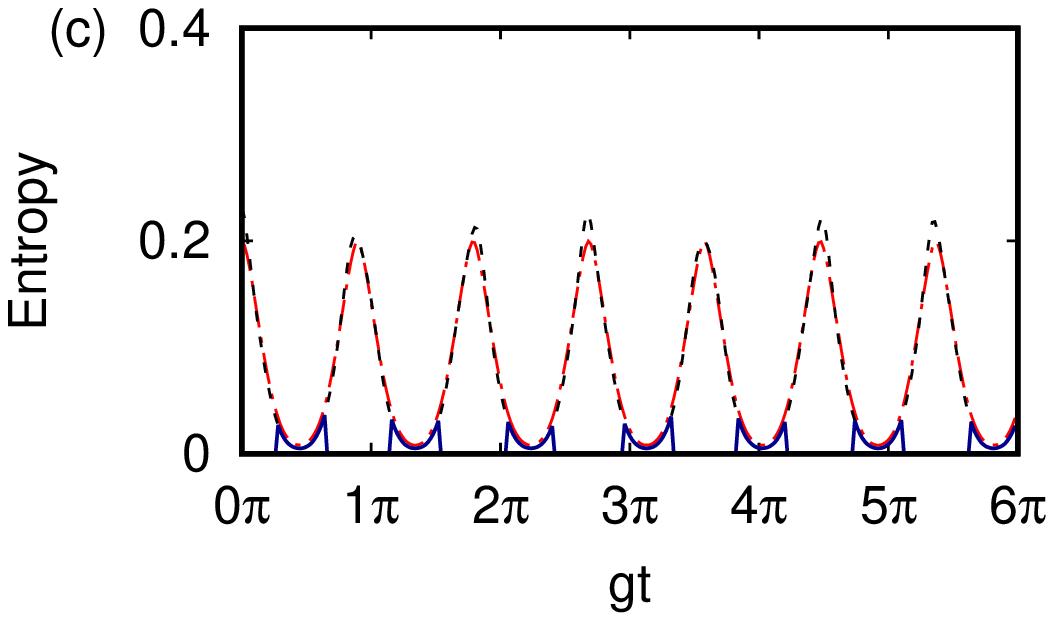}
\caption{$\xi_{\textsc{tei}}$ (black dashed curve), $\xi'_{\textsc{tei}}$ (blue solid curve) and $0.1\,\xi_{\textsc{qmi}}$ 
(red dot-dashed curve) versus $gt$  
for the atomic subsystem in the double JC model. 
 (a)  $\Delta=0$, initial state $\ket{0;0;\phi_{0}}$\,  
 (b) $\Delta = 0$, initial state $\ket{0;0;\psi_{0}}$\, 
 (c) $\Delta=1$,  initial state $\ket{0;0;\psi_{0}}$.}
\label{fig:comp_Neq1_atom}
\end{figure}

Figs. \ref{fig:comp_Neq1_atom} (a)-(c) depict plots of 
$\xi_{\textsc{tei}},  \xi'_{\textsc{tei}}$ and $\xi_{\textsc{qmi}}$ corresponding to the atomic subsystem for the same set of parameters and initial states as in Figs. \ref{fig:comp_Neq1_field}.  
In this case, although $\xi_{\textsc{tei}}$ is in good 
agreement with  $\xi_{\textsc{qmi}}$  over the time interval considered, $\xi'_{\textsc{tei}}$ is not, 
in sharp  contrast to the situation for the field subsystems. 
We note that when $\Delta=0$,  
$\xi_{\textsc{qmi}}$ returns to  its initial value of $2$ 
at the instant $gt = \pi$. 
 We will use this feature in the sequel, when we construct an equivalent circuit for the double JC model. 

\begin{figure}[h]
\centering
\includegraphics[width=\textwidth]{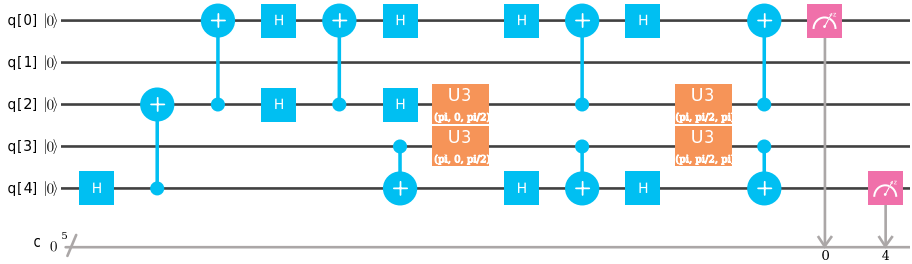}
\vspace*{1 ex}
\caption{Equivalent circuit diagram for the double JC model (created using IBM Q).}
\label{fig:circuit_dia}
\end{figure}

\begin{figure}
\includegraphics[width=0.3\textwidth]{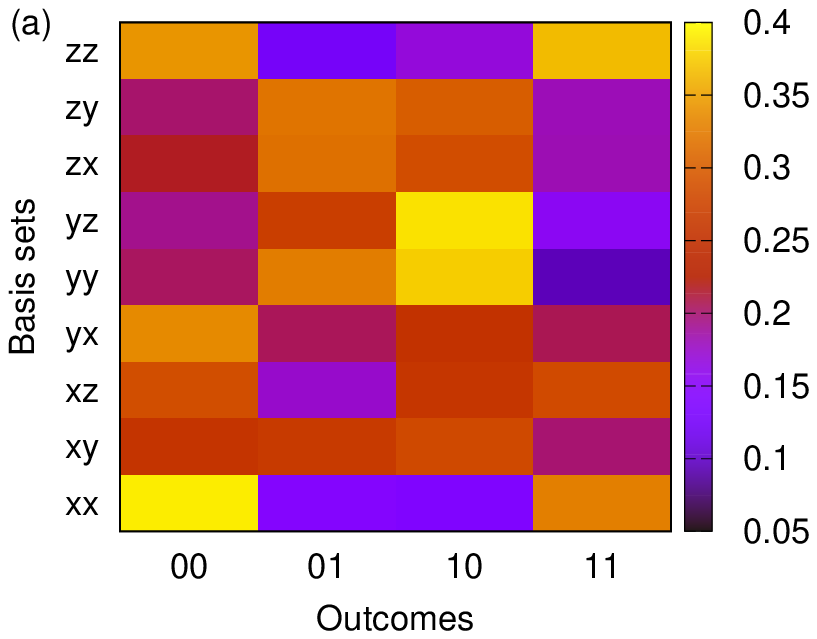}
\includegraphics[width=0.3\textwidth]{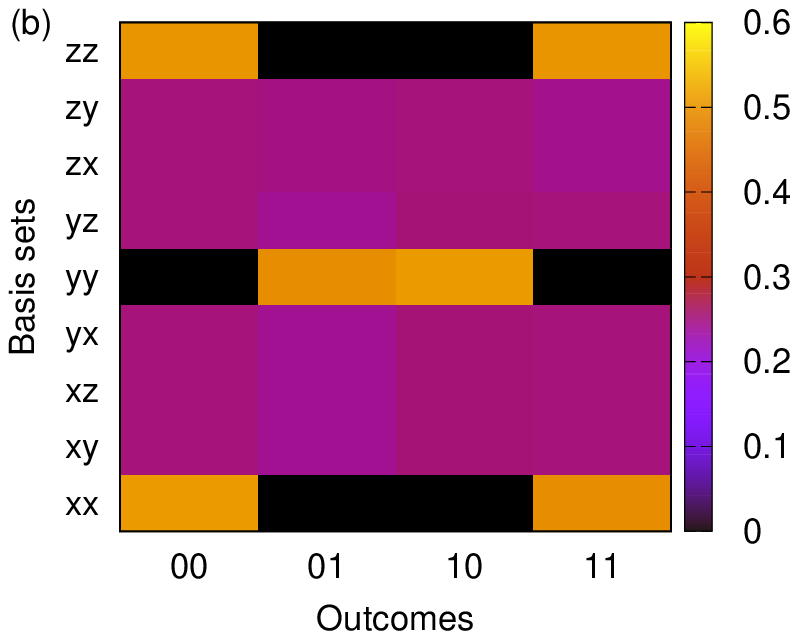}
\includegraphics[width=0.3\textwidth]{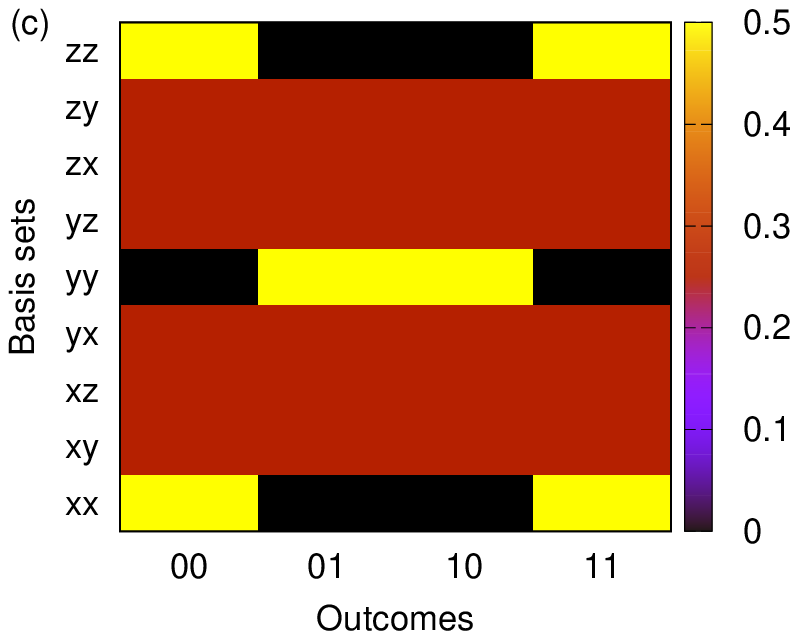}
\caption{Tomograms obtained by (a) experimental realization  
(b) simulation  of the equivalent circuit  (c) simulation of the double JC model.}
\label{fig:tomograms}
\end{figure}

The equivalent circuit from which bipartite qubit tomograms are obtained (analogous to the tomograms corresponding to the atomic subsystem of the double JC model)  is shown  in Fig. \ref{fig:circuit_dia}. We use the standard notation  
of the IBM platform \cite{ibm_main}, as the circuit has been implemented experimentally and simulated numerically using IBM Q. In the circuit,
$q[0]$ and $q[4]$ are the qubits that follow the dynamics of the atomic subsystem while $q[2]$ and $q[3]$ act as auxiliary qubits to aid the dynamics. Since  transitions between the two energy levels of either atom in the double JC model involve absorption or emission of a single photon,  each auxiliary qubit in the equivalent circuit toggles  between the qubit states $\ket{0}$ and $\ket{1}$ respectively. Here
\begin{equation}
U_{3}(\theta,\varphi,\chi)=\begin{bmatrix}
\cos\,(\theta/2) \;&\; \;-e^{i \chi}\,\sin\,(\theta/2) \\[4pt]
e^{i \varphi}\,\sin\,(\theta/2) \;
& \;\;e^{i (\chi+\varphi)}\,\cos\,(\theta/2)
\end{bmatrix},
\label{eqn:U3}
\end{equation} 
where  $0\leqslant\theta<\pi, \, 
0\leqslant \varphi<2 \pi$ and $0\leqslant\chi<2 \pi$. Each of the four qubits is initially in the qubit state $\ket{0}$. The initial entangled state between $q[0]$ and $q[4]$ (analogous to the initial state $\ket{\psi_{0}}$ of the atomic subsystem) is prepared in the circuit using an Hadamard and a controlled-NOT gate  between $q[4]$ and $q[2]$ and a SWAP gate between $q[2]$ and $q[0]$. Here, $\theta$ is analogous to $g t$ in the double JC model. We 
choose $\theta=\pi$, $\varphi=0$ and $\chi=\pi/2$. As noted earlier, the extent of entanglement is equal to its initial value ($= 2$) 
if $\theta = \pi$, and the values of $\varphi$ and $\chi$ are set for implementation of the circuit.  The matrix  
$U_{3}(\pi,\pi/2,\pi)$ which appears in the equivalent circuit is equal to $U_{3}^{\dagger}(\pi,0,\pi/2)$. 
Measurements are carried out in the $x$, $y$ and $z$ bases. A measurement in the $x$-basis is achieved by applying an Hadamard gate followed by a $z$-basis measurement. (The measurement in the $z$-basis is automatically provided by the IBM platform).  Defining the operator
\begin{equation}
S^{\dagger} = \begin{bmatrix}
1 & 0 \\
0 & -i
\end{bmatrix},
\label{eqn:sdg}
\end{equation}
measurement in the $y$-basis is achieved by applying $S^{\dagger}$, then  an Hadamard gate, and  finally a measurement in 
the $z$-basis.
Measurements in the  $x$, $y$ and $z$ bases  are needed for obtaining the spin tomogram, Fig. \ref{fig:tomograms} (a). (This is equivalent to the bipartite atomic tomogram in the double JC model, in the basis sets of $\sigma_{x}$, $\sigma_{y}$ and $\sigma_{z}$). 

These spin tomograms have been obtained  experimentally 
using the IBM superconducting circuit with appropriate Josephson junctions  (Fig. \ref{fig:tomograms} (a)), and the QASM simulator  provided by IBM which does not take into account losses at various stages of the circuit (Fig. \ref{fig:tomograms} (b)). These tomograms are compared with the atomic tomograms (Fig. \ref{fig:tomograms} (c)) of the double JC model with decoherence effects neglected.
The qualitative features are very similar in Figs. \ref{fig:tomograms} (b) and (c) as the circuit follows the dynamics of the atomic subsystem of the double JC model. As expected, Fig. \ref{fig:tomograms} (a) is distinctly different.

From these tomograms, we have calculated the corresponding tomographic entanglement indicator $\xi_{\textsc{tei}}$. The values 
obtained from the experiment, simulation and numerical analysis 
are $0.0410 \pm 0.0016$, $0.2311$ and $0.2310$,  respectively. 
In the first case, tomograms were obtained from 
six executions of the experiment (each execution is $8192$ runs over each of the 9 basis sets), and the error bar 
was calculated from the standard deviation of $\xi_{\textsc{tei}}$. Owing  to losses at various stages of the experiment, $\xi_{\textsc{tei}}$ is significantly 
smaller than the value  expected from simulation of the circuit and from  the JC model. 

It is instructive to estimate the extent of loss in state preparation 
{\em  alone}. For this purpose, an entangled state of two qubits was prepared using an Hadamard and a controlled-NOT gate, to effectively mimic $\ket{\psi_{0}}$. Tomograms were  obtained  experimentally in six trials as before, and $\xi_{\textsc{tei}}$  computed from these.  They were compared with corresponding values from numerical simulation of the entangled state and from the atomic tomogram corresponding to $\ket{\psi_{0}}$. These  values are $0.0973 \pm 0.0240$, $0.2310$ and $0.2310$ from the experiment, simulation and numerical analysis respectively. This demonstrates that substantial losses arise even in state preparation.  In order to examine the  extent to which  an increase in the number of atoms in the system increases these losses, we 
turn to  the double Tavis-Cummings (TC) model.

\subsection{The double Tavis-Cummings model}

The model comprises  four two-level atoms, $C_{1}$, $C_{2}$, $D_{1}$ and $D_{2}$, with $C_{1}$ and $C_{2}$ (respectively, $D_{1}$ and $D_{2}$)  coupled with strength $g$ to a radiation field $A$ (respectively, $B$) of frequency $\omega$.  The effective Hamiltonian (setting $\hbar=1$) is \cite{dtcm}  
\begin{align}
\nonumber H_{\textsc{DTC}}=\sum_{j=A,B} 
&\omega \,a_{j}^{\dagger} a_{j}  + \sum_{k=1}^{2}
\big\{\tfrac{1}{2}\omega_{0} \,\sigma_{\textsc{c}_{k} z} + \tfrac{1}{2}\omega_{0} \,\sigma_{\textsc{d}_{k} z} \\
&+ g ( a_{\textsc{a}}^{\dagger} \sigma_{\textsc{c}_{k} -} + a_{\textsc{a}} \sigma_{\textsc{c}_{k} +}) + g ( a_{\textsc{b}}^{\dagger} \sigma_{\textsc{d}_{k} -} + a_{\textsc{b}} \sigma_{\textsc{d}_{k} +})\big\},
\label{eqn:HDTC}
\end{align}
where the notation is self-explanatory. 
Initially, $C_{1}$ and $D_{1}$ (respectively, $C_{2}$ and $D_{2}$) are considered to be in a bipartite entangled state. This state could either be 
$\ket{\psi_{0}}$ (Eq. \ref{eqn:psi0_defn}) or 
$\ket{\phi_{0}}$ (Eq. \ref{eqn:phi0_defn}). Each field is initially in $\ket{0}$. We consider three initial states of the full system, namely, $\ket{0;0;\psi_{0};\psi_{0}}$, 
$\ket{0;0;\phi_{0};\phi_{0}}$ and 
$\ket{0;0;\psi_{0};\phi_{0}}$. The notation 
 $\ket{0;0;\psi_{0};\phi_{0}}$ indicates, 
 for instance, that  
 $A$ and $B$ are in the state $\ket{0}$, the bipartite subsystem $(C_{1}, D_{1})$ is in the state 
 $\ket{\psi_{0}}$,  and the bipartite 
 subsystem $(C_{2}, D_{2})$ is in 
 the state $\ket{\phi_{0}}$. For brevity,  
 we refer to  the bipartite atomic subsystems 
 $(C_{1}, C_{2})$  and 
 $(D_{1}, D_{2})$ as subsystems $C$ and $D$, 
 respectively.  

\begin{figure}
\includegraphics[width=0.8\textwidth]{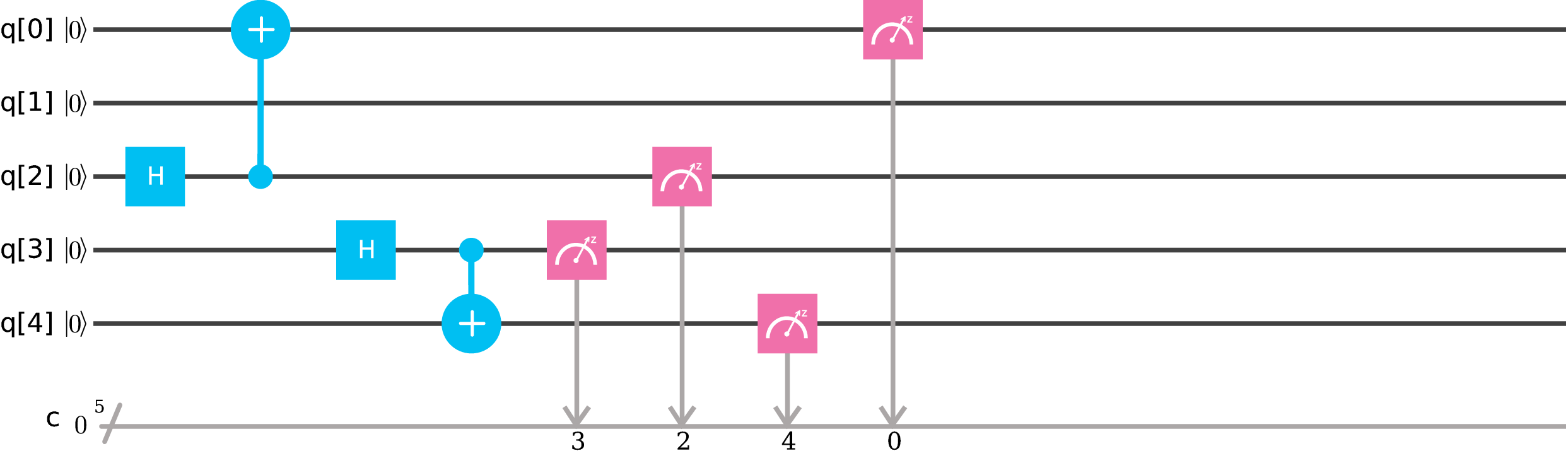}
\caption{Circuit for preparing an entangled state analogous to 
$\ket{\psi_{0};\psi_{0}}$ (created using IBM Q). }
\label{fig:DTC_circuit}
\end{figure}

An equivalent circuit  for the double TC model  will require 4 qubits to mimic the four  two-level atoms together with a minimum of 4 auxiliary qubits to aid the dynamics. In order to  assess the extent of losses in  state preparation  {\em alone}, 4 qubits were prepared in a pairwise entangled state (analogous to the initial state $\ket{\psi_{0};\psi_{0}}$ of the atomic subsystem $(C,D)$) using $2$ Hadamard and $2$ controlled-NOT gates (Fig. \ref{fig:DTC_circuit}). Here qubits $q[2]$ and $q[3]$ are entangled with qubits $q[0]$ and 
$q[4]$ respectively.  As in the earlier case, tomograms have been  obtained (a) experimentally using the IBM quantum computer, (b) from the QASM simulator, and (c) from the corresponding atomic tomogram for 
$\ket{\psi_{0};\psi_{0}}$. Note that the pair ($q[2]$,$q[3]$) is analogous to subsystem $C$,  and ($q[0]$,$q[4]$) is analogous to $D$. We have calculated the extent of entanglement 
$\xi_{\textsc{tei}}$ between  the two $2$-qubit subsystems. The numerical values obtained 
from (a), (b) and (c)
are $0.2528$, $0.4761$ and $0.4621$, respectively. As 4 qubits are involved in this circuit, the number of possible outcomes is 16. (Recall that the number of outcomes in the earlier case was 4.) The maximum number of experimental runs possible in both cases is 8192\cite{ibm_main}. Hence, the experimental losses,  as well as the difference between the simulated and the numerically obtained values,  are higher than those obtained for  the double JC model. We therefore proceed to investigate the entanglement dynamics of the double TC model numerically in the absence of losses.

\begin{figure}[h]
\centering
\includegraphics[width=0.32\textwidth]{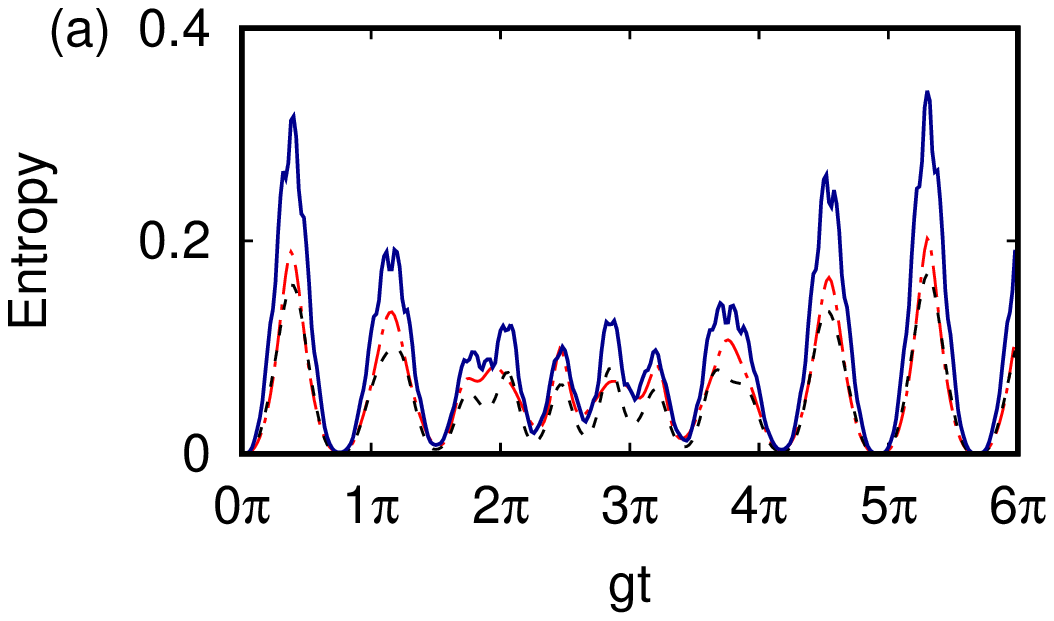}
\includegraphics[width=0.32\textwidth]{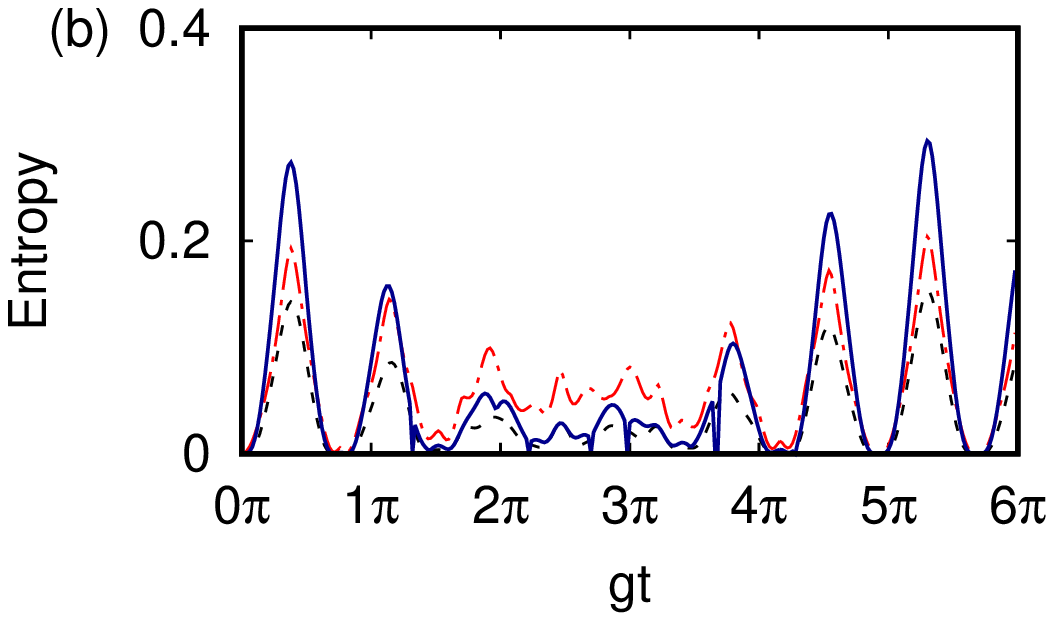}
\includegraphics[width=0.32\textwidth]{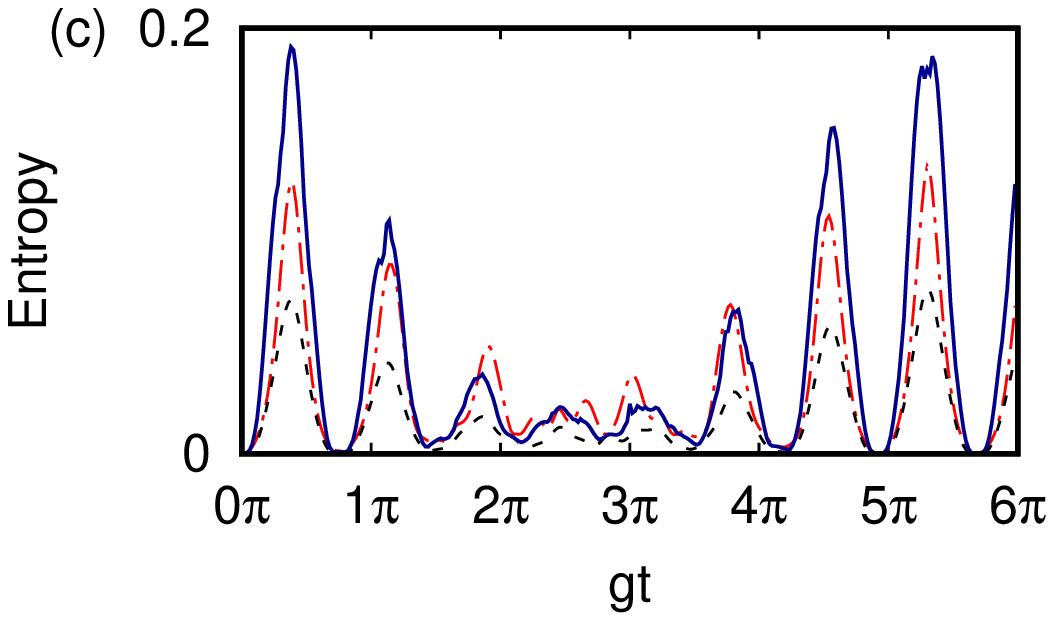}
\caption{$\xi_{\textsc{tei}}$ (black dashed curve), $\xi'_{\textsc{tei}}$ (blue solid curve) and $\xi_{\textsc{qmi}}/10$ (red dot-dashed curve) between the two radiation fields $A$ and $B$ versus time $g t$ with $\Delta=0$. The fields 
are in the initial state in $\ket{0;0}$, while the atoms are 
in (a) $\ket{\psi_{0}; \psi_{0}}$\, (b) $\ket{\phi_{0};
\phi_{0}}$\, (c) $\ket{\psi_{0};\phi_{0}}$.}
\label{fig:comp_Neq2_field}
\end{figure}

\begin{figure}[h]
\centering
\includegraphics[width=0.32\textwidth]{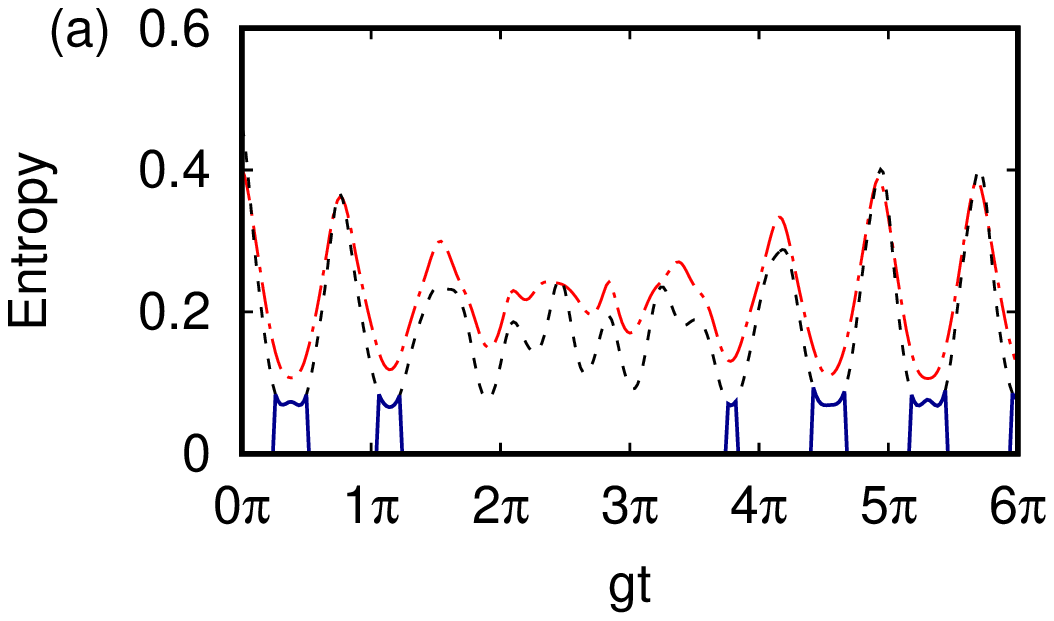}
\includegraphics[width=0.32\textwidth]{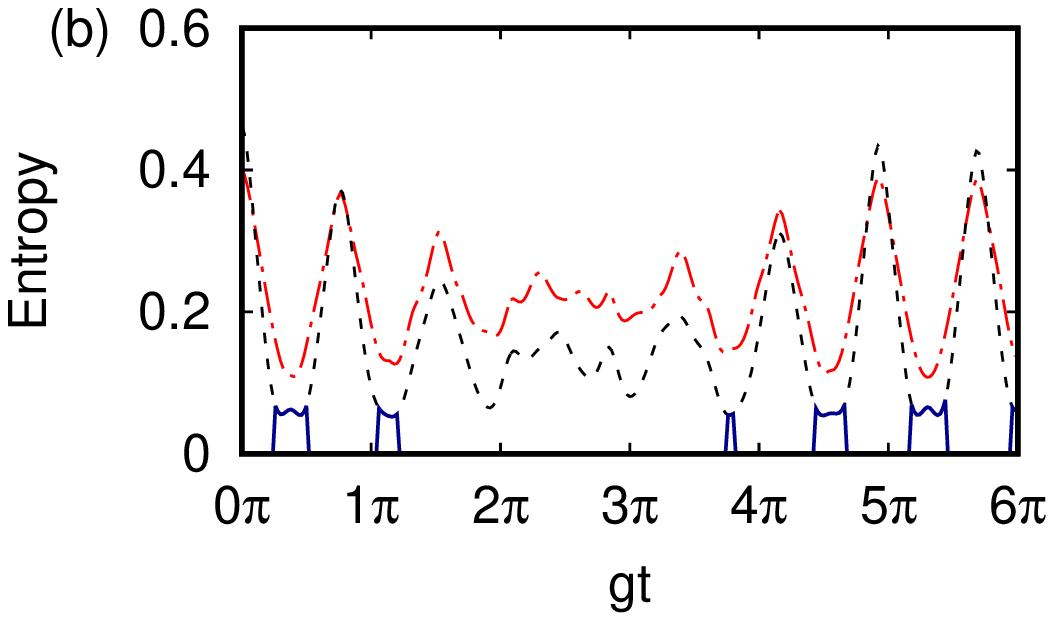}
\includegraphics[width=0.32\textwidth]{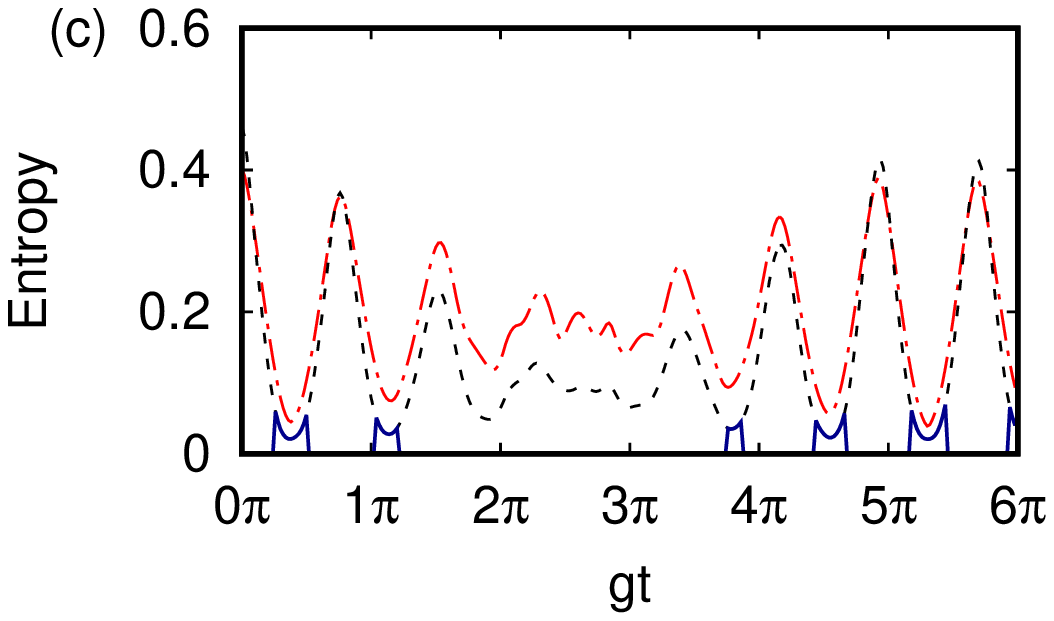}
\caption{$\xi_{\textsc{tei}}$ (black dashed curve), $\xi'_{\textsc{tei}}$ (blue solid curve) and $\xi_{\textsc{qmi}}/10$ (red dot-dashed curve) between the two atomic subsystems $C$ and $D$ versus  $g t$ with $\Delta=0$. The initial states 
of the fields and atoms are the same as in 
Figs. \ref{fig:comp_Neq2_field} (a)-(c).}
\label{fig:comp_Neq2_det0_atom}
\end{figure}

We have investigated the  entanglement between the field subsystems  $A$ and $B$,  and between  the atomic subsystems $C$ and $D$, as the system evolves in time.  For this purpose, we have numerically generated the states of the full system  and the corresponding  tomograms during temporal evolution, at 300  instants separated by a time step 0.02 in units of $\pi/g$.  From these, we have obtained the two entanglement indicators $\xi_{\textsc{tei}}$ and 
$\xi'_{\textsc{tei}}$ as  functions of time. 
These indicators and the corresponding $\xi_{\textsc{qmi}}$ between $A$ and $B$ are plotted in Figs. \ref{fig:comp_Neq2_field} (a)-(c). (Here, the detuning 
parameter may be set equal to  zero, 
without loss of generality.) $\xi_{\textsc{tei}}$, 
$\xi'_{\textsc{tei}}$ and $\xi_{\textsc{qmi}}$ corresponding to entanglement between $C$ and $D$ are plotted in Figs. \ref{fig:comp_Neq2_det0_atom} (a)-(c).
 
The inferences drawn from the double JC model 
are seen to hold 
good in this case too: namely, that both the indicators effectively mimic $\xi_{\textsc{qmi}}$ for the field 
subsystem, while  $\xi'_{\textsc{tei}}$ does not 
reflect  $\xi_{\textsc{qmi}}$ for the atomic subsystem.

\section{Concluding remarks}
\label{conclremarks}

We have compared an entanglement indicator 
$\xi_{\textsc{tei}}$ obtained directly from tomograms,
 and an approximation to it ($\xi'_{\textsc{tei}}$),  
 with the quantum mutual information $\xi_{\textsc{qmi}}$, 
 in the case of the double JC and the double TC models. In both models, the approximation is satisfactory for the field subsystem, but  not for the atomic subsystem. 
 $\xi_{\textsc{tei}}$, however,  is found  to be a good estimate in both models and for both subsystems. This would imply that a good entanglement indicator could be obtained directly from tomograms, circumventing error-prone 
 and lengthy  procedures of state reconstruction in multipartite hybrid systems involving field-atom interactions. An equivalent circuit for the double JC model was both experimentally run and numerically simulated to obtain entanglement indicators. This facilitates the estimation of experimental losses and establishes that the results from the IBM simulator agree well with numerical simulation of the double JC model. By constructing equivalent circuits for state preparation in both models, we have shown  that the  difference in the values of the entanglement indicator obtained experimentally and numerically increases significantly  with an increase in 
 the number of atoms. 

\begin{acknowledgements}
We acknowledge use of the IBM Q for this work. The views expressed are those of the authors and do not reflect the official policy or position of IBM or the IBM Q team. We have also used the Department Computing Facility, Department of Physics, IIT Madras.
\end{acknowledgements}

\bibliographystyle{spphys}      
\bibliography{references}   

\end{document}